\documentclass{Interspeech2024}
\usepackage{hyperref}

\interspeechcameraready

\title{FastAST: Accelerating Audio Spectrogram Transformer via Token Merging and Cross-Model Knowledge Distillation}
\name[affiliation={\dagger\ast}]{Swarup Ranjan}{Behera}
\name[affiliation={\dagger}]{Abhishek}{Dhiman}
\name[affiliation={}]{Karthik}{Gowda}
\name[affiliation={}]{Aalekhya Satya}{Narayani}

\address{Reliance Jio AICoE, Hyderabad, India}
\email{\{swarup.behera,abhishek1.dhiman\}@ril.com, \{karthik4.gowda,Aalekhya.Narayani\}@zmail.ril.com}

\keywords{audio spectrogram transformer, token merging, cross model knowledge distillation}

\begin{document}

\maketitle

\let\thefootnote\relax\footnotetext{
    \hspace{0em}\parbox{\textwidth}{
    \textsuperscript{*} Corresponding Author\\
    \textsuperscript{†} Authors contributed equally as first authors}
}

\begin{abstract}
Audio classification models, particularly the Audio Spectrogram Transformer (AST), play a crucial role in efficient audio analysis. However, optimizing their efficiency without compromising accuracy remains a challenge. In this paper, we introduce FastAST, a framework that integrates Token Merging (ToMe) into the AST framework. FastAST enhances inference speed without requiring extensive retraining by merging similar tokens in audio spectrograms. Furthermore, during training, FastAST brings about significant speed improvements. The experiments indicate that FastAST can increase audio classification throughput with minimal impact on accuracy. To mitigate the accuracy impact, we integrate Cross-Model Knowledge Distillation (CMKD) into the FastAST framework. Integrating ToMe and CMKD into AST results in improved accuracy compared to AST while maintaining faster inference speeds. FastAST represents a step towards real-time, resource-efficient audio analysis.
\end{abstract}

\section{Introduction}

    In recent years, the field of audio classification has witnessed significant advancements, primarily driven by the integration of deep learning techniques. The Audio Spectrogram Transformer (AST)~\cite{AST1,AST2,AST3}, specifically designed for audio classification, has emerged as a powerful tool. Leveraging self-attention mechanisms of transformer~\cite{Trans}, AST processes audio data efficiently, bypassing traditional convolutional layers~\cite{CNN1,CNN2,CNN3} and achieving remarkable results across diverse audio classification tasks. Despite its success, the computational demands of AST, $\mathcal{O}(n^2)$ complexity, present a considerable obstacle, limiting its real-time and resource-efficient deployment in practical applications. The abundance of tokens in the input spectrogram, representing various time-frequency audio signal representations, incurs substantial computational overhead during inference.

    Existing methods~\cite{xFormers} have successfully sped up transformers, but they do not inherently reduce the overall computational workload, as they still evaluate the transformer for each token. Spectrograms, characterized by inherent redundancy, lead to inefficient and resource-intensive computing operations for each token. Advancements in token reduction techniques, as showcased in \textit{token pruning}~\cite{TP1,TP2,TP3} and \textit{token merging}~\cite{TOME,TOME2}, have proven effective in eliminating redundant tokens in transformers. This accelerates the evaluation process with only a slight reduction in accuracy. While most of these approaches typically require model re-training, incurring significant costs, Token Merging (ToMe)~\cite{TOME} distinguishes itself by not necessitating any additional training.

    In this paper, we introduce FastAST, a novel approach meticulously designed to enhance the efficiency of AST. FastAST incorporates ToMe~\cite{TOME}, initially developed to accelerate Vision Transformer (ViT)~\cite{VIT} models in computer vision, offering a promising solution to the computational challenges faced by AST. FastAST enhances both inference and training speeds by merging similar tokens in audio spectrograms. This optimization allows FastAST to increase audio classification throughput with minimal impact on accuracy. 
    
    To enhance the accuracy of FastAST while maintaining its efficiency gains through token merging, we integrate Cross-Model Knowledge Distillation (CMKD)~\cite{CMKD} into the FastAST. CMKD leverages the remarkable synergy observed between different model architectures, such as Convolutional Neural Networks (CNNs)~\cite{CNN1,CNN2,CNN3} and ASTs~\cite{AST1,AST2,AST3}. In particular, the CNN/AST CMKD method has demonstrated exceptional performance improvements on various audio classification benchmarks, surpassing the capabilities of individual models. This approach capitalizes on the unique strengths of each model type: CNNs excel in capturing spatial locality and translation equivariance, while AST offer flexibility and data-driven learning. By distilling knowledge from a more accurate reference model, CMKD empowers FastAST to surpass AST in accuracy, mitigating the impact of token merging while preserving faster inference. This integration not only enriches FastAST's performance but also underscores the versatility and effectiveness of CMKD techniques in advancing audio classification capabilities. 

    The contribution of this paper is threefold:

    \begin{itemize}
        \item Demonstrating how FastAST seamlessly integrates ToMe with AST, enabling significant speed improvements without the need for extensive retraining. Through empirical evidence, we show that FastAST can double audio classification throughput with only marginal accuracy trade-offs.
        \item Exploring the utilization of ToMe during the training of FastAST, revealing additional speed improvements and emphasizing the synergistic relationship between ToMe and AST.
        \item Integrating CMKD into FastAST to address the accuracy impact observed with token merging. By leveraging the remarkable synergy between different model architectures, such as CNN and AST, CMKD empowers FastAST to surpass AST in accuracy while preserving faster inference speeds.
    \end{itemize}

   Collectively, these contributions propel the field of audio classification, addressing key challenges and advancing the state-of-the-art. The code can be accessed at \url{https://github.com/swarupbehera/FastAST}.

   \begin{figure*}[!tbh]
     \centering
       \includegraphics[width=\linewidth]{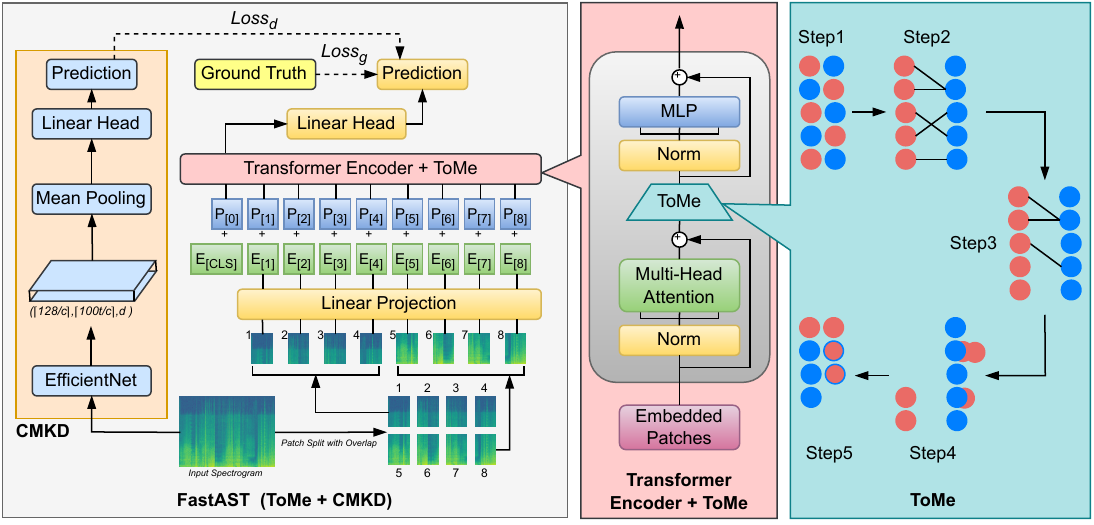}
      \caption{Visualization of the proposed framework, \textbf{FastAST}, illustrating the integration of Token Merging (ToMe) and Cross-Model Knowledge Distillation (CMKD) into AST in the LHS box. The middle box demonstrates the incorporation of ToMe in the transformer encoder of AST, while the RHS box outlines the steps of the ToMe process. On the LHS box, the CMKD block depicts the process of cross-model knowledge distillation where a CNN, EfficientNet, serves as the teacher model.}
      \label{fig:archi}
    \end{figure*}

    \section{FastAST}  \label{sec:fastast}
    In this section, we examine different elements of the proposed framework, \textbf{FastAST}, illustrated in Figure \ref{fig:archi}. Initially, we explore the AST architecture, laying the foundation for the FastAST approach. Following that, we investigate the intricacies of the ToMe technique and its incorporation into AST, which forms a core aspect of FastAST. Finally, we present CMKD and its integration, with the goal of improving FastAST's accuracy.
    
    \subsection{Audio Spectrogram Transformer (AST)}
    In this research, we utilize the original AST framework~\cite{AST1}, illustrated in the LHS box of Figure \ref{fig:archi}. The input audio waveform, spanning a duration of $t$ seconds, is transformed into a $128  \times  100t$ spectrogram, following a similar process as in CNN models. Subsequently, the spectrogram is divided into a sequence of $N$ $16 \times 16$ patches with a temporal and frequency dimension overlap of 6, where $N = 12 \lceil (100t - 16)/10 \rceil$ represents the patch count, serving as the effective input sequence length for the Transformer. Each $16 \times 16$ patch is flattened into a 1D patch embedding of size $d$ using a linear projection layer, referred to as the patch embedding layer. To capture the spatial structure of the 2D audio spectrogram, a trainable positional embedding of size $d$ is added to each patch embedding. Following this, a $[CLS]$ token is added to the sequence, and the resulting sequence is inputted into a standard Transformer encoder. The output of the Transformer encoder's $[CLS]$ token serves as the audio spectrogram representation, which is then utilized for classification using a linear layer with sigmoid activation for multi-label classification or softmax for single-label classification.
    
    \subsection{Token Merging (ToMe)}
    Token Merging, also referred to as ToMe~\cite{TOME}, systematically reduces the number of tokens in a transformer by merging `r' tokens in each block. In FastAST (LHS box of Figure~\ref{fig:archi}), ToMe is incorporated in each blocks of Transformer encoder. As depicted in middle box of Figure~\ref{fig:archi}, we integrate ToMe between the attention and MLP branches of every transformer block. This strategic placement facilitates the propagation of information from tokens slated for merging, and leverages features within attention to determine the merging strategy, leading to enhanced accuracy. 
    
    Please refer to the RHS box of Figure~\ref{fig:archi} for the sequence of token merging steps. ToMe begins by partitioning the tokens into two sets, A and B, of approximately equal size. Next, an edge is drawn from each token in set A to its most similar token in set B. After that, the `r' most similar edges are retained. Tokens that are still connected are then merged, often by averaging their features. Finally, the two sets are concatenated back together, thereby reducing the total number of tokens by `r'. This optimization contributes to faster execution of subsequent blocks in the model. Throughout the $L$ blocks in the network, we progressively merge $rL$ tokens. Adjusting the value of `r' introduces a trade-off between speed and accuracy, where reducing the number of tokens (high value of r) results in decreased accuracy but increased throughput.
    
    \subsection{Cross-Model Knowledge Distillation (CMKD)} \label{sec:cmkd}
    To boost FastAST's accuracy while maintaining its efficiency gains from token merging, we integrate Cross-Model Knowledge Distillation (CMKD) into FastAST training. CMKD harnesses the synergy between CNNs and ASTs, leveraging their unique strengths: CNNs for spatial locality and ASTs for flexibility. By distilling knowledge from a more accurate model, CMKD enhances FastAST's accuracy, offsetting the impact of token merging while preserving faster inference.

    In line with the CMKD\cite{CMKD} paper, we adopt the original knowledge distillation setup~\cite{KD1} with consistent teaching~\cite{KD2}. First the teacher model is trained. Then, when training the student model, the same augmentations are used on the input audio spectrogram for both models, ensuring consistency in teaching. The following loss function is employed for training the student model:
    \begin{equation}
        \text{Loss} = \lambda \text{Loss}_g(\psi(\mathbf{Z}_s), y) + (1-\lambda) \text{Loss}_d(\psi(\mathbf{Z}_s), \psi(\mathbf{Z}_{t}/\tau))
    \end{equation}
    where \( \lambda \) is the balancing coefficient; \( \text{Loss}_\text{g} \) and \( \text{Loss}_\text{d} \) are the ground truth and distillation losses, respectively; \( \psi \) represents the activation function; \( \mathbf{Z}_\text{s} \) and \( \mathbf{Z}_\text{t} \) are the logits of the student and teacher model, respectively; $y$ is the true label; \( \tau \) is the temperature. The teacher model is frozen during the student model training.

    Drawing inspiration from the CMKD paper, we conduct two cross-model knowledge distillations, employing a CNN-based model, EfficientNet-B2 \cite{efficientnetb2}, and a Transformer-based model, PaSST \cite{passt}, as the teacher models. The direction of knowledge distillation is denoted as teacher→student. \textbf{(i) EfficientNet-B2→AST:} First, an EfficientNet-B2 model is trained and subsequently it is employed as the teacher to train an AST model as the student. \textbf{(ii) PaSST→AST:} Initially, a PaSST model is trained and subsequently it is utilized as the teacher to train an AST model as the student.

    \section{Experimental Results and Analysis}
    \label{sec:ExperimentalResultsandAnalysis}

    We conduct a comprehensive evaluation of FastAST on audio classification tasks using three distinct datasets: Environmental Sound Classification (ESC-50) \cite{ESC}, Speech Commands V2 \cite{SPEECHCOMMAND}, and AudioSet \cite{AUDIOSET}. ESC-50 consists of 2,000 5-second audio recordings thoughtfully organized into 50 classes, designed for the detailed classification of various environmental sounds. Speech Commands V2 encompasses 105,829 1-second recordings containing 35 common speech commands. This dataset provides dedicated sets for training, validation, and testing, comprising 84,843, 9,981, and 11,005 samples, respectively. It is tailored for speech command recognition tasks, focusing on a 35-class classification challenge. AudioSet comprises 2 million 10-second audio clips from YouTube, categorized into 527 sound labels. The dataset includes balanced training, full training, and evaluation sets with 22k, 2M, and 20k samples, respectively. For our experiments, we select the balanced training set of the AudioSet dataset, denoted as `Balanced Audioset.'
    
    We perform two sets of experiments. Firstly, we showcase the effectiveness of incorporating ToMe into AST (FastAST (ToMe)), followed by integrating both ToMe and CMKD (FastAST (ToMe+CMKD)).

    \subsection{FastAST (ToMe)}

        \begin{table}[!t]
      \caption{FastAST (ToMe) results on ESC-50, Speech Commands V2, and  Balanced Audioset dataset. ToMe applied only during inference (FastAST (Inf)). ToMe applied during both training and inference (FastAST (Train+Inf)). Reduction factor (r). The metrices reported are accuracy (Accuracy) for ESC-50 and Speech Commands V2 and  mean Average Precision (mAP) for Balanced Audioset. Drop in Accuracy or mAP (Drop). Sample per second (S/s).}
      \label{tab:result1}
      \centering
      \begin{tabular}{c|ccc|ccc}
      \toprule
    \multicolumn{1}{l}{} & \multicolumn{3}{c}{\textbf{FastAST (Inf)}} & \multicolumn{3}{c}{\textbf{FastAST (Train + Inf)}} \\
    \midrule
    \multicolumn{7}{c}{\textbf{ESC-50}}                                                                                                   \\
    \midrule
    \textbf{r}           & \textbf{Accuracy}   & \textbf{Drop}   & \textbf{S/s}  & \textbf{Accuracy}       & \textbf{Drop}      & \textbf{S/s}      \\
    \midrule
    0                    & 95.6         & 0           & 65         & 94.7            & 0               & 65             \\
5                    & 95.2         & -0.4        & 75         & 94.5            & -0.2            & 75             \\
10                   & 95.1         & -0.5        & 85         & 94.4            & -0.3            & 85             \\
15                   & 94.9         & -0.7        & 95         & 94.3            & -0.4            & 95             \\
20                   & 94.2         & -1.4        & 110        & 94.3            & -0.4            & 110            \\
25                   & 93.1         & -2.5        & 120        & 93.7            & -1              & 120            \\
30                   & 92.4         & -3.2        & 125        & 93.4            & -1.3            & 125            \\
35                   & 91.8         & -3.8        & 140        & 93.1            & -1.6            & 140            \\
40                   & 90.2         & -5.4        & 150        & 92.9            & -1.8            & 150            \\
    \midrule
    \multicolumn{7}{c}{\textbf{Speech Commands V2}}                                                                                       \\
    \midrule
    \textbf{r}           & \textbf{Accuracy}   & \textbf{Drop}   & \textbf{S/s}  & \textbf{Accuracy}       & \textbf{Drop}      & \textbf{S/s}      \\
    \midrule
    0                    & 98.1         & 0           & 340        & 97.8            & 0               & 340            \\
5                    & 98.1         & 0           & 380        & 97.7            & -0.1            & 380            \\
10                   & 97.9         & -0.2        & 410        & 97.5            & -0.3            & 410            \\
15                   & 97.4         & -0.7        & 455        & 97.5            & -0.3            & 455            \\
20                   & 97.2         & -0.9        & 475        & 97.4            & -0.4            & 475            \\
25                   & 96.3         & -1.8        & 490        & 97.1            & -0.7            & 490            \\
30                   & 95.2         & -2.9        & 512        & 96.8            & -1              & 512            \\
35                   & 94.5         & -3.6        & 540        & 96.6            & -1.2            & 540            \\
40                   & 93.3         & -4.8        & 580        & 96.4            & -1.4            & 580            \\
    \midrule
    \multicolumn{7}{c}{\textbf{Balanced Audioset}}                                                                                       \\
    \midrule
    \textbf{r}           & \textbf{mAP}   & \textbf{Drop}   & \textbf{S/s}  & \textbf{mAP}       & \textbf{Drop}      & \textbf{S/s}      \\
    \midrule
    0                    & 38.4         & 0           & 55         & 38.2            & 0               & 55             \\
5                    & 38.2         & -0.2        & 59         & 38.1            & -0.1            & 59             \\
10                   & 37.7         & -0.7        & 60         & 37.9            & -0.3            & 60             \\
15                   & 37.7         & -0.7        & 62         & 37.8            & -0.4            & 62             \\
20                   & 37.5         & -0.9        & 65         & 37.6            & -0.6            & 65             \\
25                   & 37.9         & -0.5        & 68         & 37.5            & -0.7            & 68             \\
30                   & 37.2         & -1.2        & 70         & 37.3            & -0.9            & 70             \\
35                   & 38.1         & -0.3        & 74         & 37.1            & -1.1            & 74             \\
40                   & 36           & -2.4        & 80         & 36.9            & -1.3            & 80                  \\
    \bottomrule
    \end{tabular}
        \end{table} 

    \begin{table*}[!t]
      \caption{FastAST with and without knowledge distillation results on Balanced Audioset dataset. FastAST without knowledge distillation (FastAST). FastAST with knowledge distillation where a transformer based model, Passt, is the teacher model (FastAST + KD (Passt)). FastAST with knowledge distillation where a CNN based model is the teacher model (FastAST + KD (CNN)). Reduction factor (r). The metrices reported are accuracy (Acc) and  mean Average Precision (mAP). Drop in Acc or mAP (Drop). Sample per second (S/s).}
      \label{tab:result2}
      \centering
    \begin{tabular}{c|ccccc|ccccc|ccccc}
    \toprule
        \multicolumn{1}{l}{} & \multicolumn{5}{c}{\textbf{FastAST}}                                                                 & \multicolumn{5}{c}{\textbf{FastAST + KD (Passt)}}                                                            & \multicolumn{5}{c}{\textbf{FastAST + KD (CNN)}}                                                              \\
        \midrule
        \textbf{r}                    & \textbf{Acc}   & {\textbf{Drop}} & \textbf{mAP}   & {\textbf{Drop}} & \textbf{S/s} & \textbf{Acc}   & {\textbf{Drop}} & \textbf{mAP}   & {\textbf{Drop}} & \textbf{S/s} & \textbf{Acc}   & {\textbf{Drop}} & \textbf{mAP}   & {\textbf{Drop}} & \textbf{S/s} \\
        \midrule
        0                    & 38.4 & 0         & 37.3 & 0         & 55        & 39.2 & 0         & 42.8 & 0         & 55        & 41.4 & 0         & 43.2 & 0         & 56        \\
5                    & 38.2 & -0.2      & 38.7 & 1.4       & 59        & 38.3 & -0.9      & 42.4 & -0.4      & 57        & 39.9 & -1.5      & 44.6 & 1.4       & 58        \\
10                   & 37.7 & -0.7      & 37.2 & -0.1      & 60        & 37.4 & -1.8      & 40.1 & -2.7      & 59        & 38.6 & -2.8      & 43.1 & -0.1      & 60        \\
15                   & 37.7 & -0.7      & 37.1 & -0.2      & 62        & 37.1 & -2.1      & 40.6 & -2.2      & 62        & 39   & -2.4      & 43.3 & 0.1       & 62        \\
20                   & 37.5 & -0.9      & 38.4 & 1.1       & 65        & 37.8 & -1.4      & 42.2 & -0.6      & 65        & 40.8 & -0.6      & 43.7 & 0.5       & 64        \\
25                   & 37.9 & -0.5      & 37   & -0.3      & 68        & 34.8 & -4.4      & 40.8 & -2        & 67        & 37.1 & -4.3      & 42.1 & -1.1      & 68        \\
30                   & 37.2 & -1.2      & 36.4 & -0.9      & 70        & 35.9 & -3.3      & 40.3 & -2.5      & 72        & 36.7 & -4.7      & 41.3 & -1.9      & 73        \\
35                   & 38.1 & -0.3      & 36.5 & -0.8      & 74        & 35.6 & -3.6      & 38.7 & -4.1      & 76        & 36.1 & -5.3      & 38.7 & -4.5      & 77        \\
40                   & 36   & -2.4      & 37.1 & -0.2      & 80        & 35.9 & -3.3      & 37.4 & -5.4      & 80        & 36.9 & -4.5      & 39.4 & -3.8      & 80    \\

        \bottomrule
        \end{tabular}
            \end{table*}
    
    In Table \ref{tab:result1}, we present the experimental results of FastAST using only ToMe across these three datasets. We vary the reduction factor ($r$), where ToMe merges `r' tokens in each transformer block, from 0 to 40 and observe accuracy for ESC-50 and Speech Commands V2, and mean Average Precision for Balanced Audioset to address class imbalance. We conduct two sets of experiments. Initially, we apply ToMe solely during inference (FastAST (Inf)). Subsequently, we apply ToMe during both training and inference (FastAST (Train+Inf)).

    \subsubsection{FastAST (Inf)}

    In Table \ref{tab:result1}, we present the results of applying ToMe solely during inference on the LHS. In the ESC-50 dataset, the baseline AST model ($r=0$) achieves an accuracy of 95.6\% with a throughput of 65 samples per second. Increasing the reduction factor ($r$) from 5 to 40 leads to a gradual decline in accuracy, ranging from 95.2\% to 90.2\%. This reduction in accuracy corresponds proportionally to the increase in $r$, resulting in dropout values ranging from -0.4\% to -5.4\% compared to the baseline. Despite the decline in accuracy, FastAST demonstrates significant throughput improvements, ranging from 75 S/s to 150 S/s for reduction factors from $r=5$ to $r=40$.\\
    
    Across both the Speech Commands V2 dataset and the Balanced Audioset dataset, similar trends are observed. The baseline AST model achieves high accuracy and throughput, with the accuracy gradually decreasing as the reduction factor ($r$) increases from 5 to 40. Despite this decrease in accuracy, FastAST consistently demonstrates notable improvements in throughput across both datasets. For the Speech Commands V2 dataset, throughput ranges from 380 S/s to 580 S/s for reduction factors from $r=5$ to $r=40$. Similarly, for the Balanced Audioset dataset, throughput ranges from 59 S/s to 80 S/s for the same reduction factors.

    \subsubsection{FastAST (Train + Inf)}
    In Table \ref{tab:result1}, we present the results of applying ToMe during both training and inference in RHS. In the comparison between FastAST (Inf) and FastAST (Train+Inf), an interesting trend emerges. At a reduction factor ($r$) of 0, FastAST (Inf) demonstrates superiority over FastAST (Train+Inf), exhibiting higher accuracy. However, as the reduction factor increases, particularly at $r>=30$, FastAST (Train+Inf) begins to outperform FastAST (Inf) in terms of accuracy. This trend holds true across all three datasets, further emphasizing the consistency of the observed phenomenon. This observation underscores the nuanced performance dynamics between the two approaches, wherein their relative effectiveness varies with the applied reduction factor. Such insights shed light on the nuanced interplay between inference strategies and training methodologies in the context of FastAST.

    In summary, FastAST significantly accelerates the inference speed of AST models with minimal impact on classification accuracy across diverse audio datasets. Adjusting the reduction factor ($r$) allows for balancing speed and accuracy, though its impact varies based on dataset complexity. FastAST's performance hinges on data quality and hardware constraints, requiring careful adaptation for optimal results.

    \subsection{FastAST (ToMe+CMKD)}
    
    In the previous section, we noted a slight decrease in accuracy when ToMe was applied to AST. To address this, we integrate CMKD into FastAST training, aiming to improve accuracy while maintaining efficiency. Given that FastAST(ToMe) exhibits the lowest accuracy or mAP on the Balanced Audioset dataset, we dive deeper into it in this experiment. As described in Section~\ref{sec:cmkd}, we conduct two experiments: (i) EfficientNet-B2→AST and (ii) PaSST→AST, following the distillation setup and hyperparameters outlined in the CMKD paper. Optimal results are achieved with $\lambda = 0.1$ and $\tau = 1$.
    
    In Table \ref{tab:result2}, we present the experimental results. The first section of the table showcases the performance of FastAST without knowledge distillation. Subsequently, the following sections, labeled as FastAST + KD (PaSST) and FastAST + KD (CNN), integrate knowledge distillation using PaSST and an EfficientNet-B2 model as the teacher models, respectively. Comparing the results, FastAST achieves an initial accuracy of 38.4\% and mAP of 37.3\% at a throughput of 55 S/s. FastAST + KD (Passt) shows slight improvement, with an initial accuracy of 39.2\% and mAP of 42.8\%, maintaining the same throughput. FastAST + KD (CNN) surpasses both, starting with an accuracy of 41.4\% and mAP of 43.2\%, albeit with a slightly higher throughput of 56 S/s. Across different reduction factors ($r$), FastAST + KD (CNN) consistently outperforms, with the highest mAP values. Even at the highest reduction factor ($r=40$), FastAST + KD (CNN) achieves a superior mAP compared to FastAST. Hence, employing knowledge distillation with a CNN-based model as the teacher yields the most promising results for enhancing audio classification.

\section{Conclusion}
    \label{sec:Conclusion}
    In conclusion, FastAST represents a significant advancement in the realm of audio classification, offering a balance between inference speed and classification accuracy. By integrating ToMe and CMKD into the AST framework, FastAST achieves notable improvements in throughput without compromising accuracy. Our experiments demonstrate that FastAST enhances inference speed while maintaining accuracy across various datasets and reduction factors. Notably, employing CMKD with a CNN-based model as the teacher yields the most promising results. These findings underscore the potential of FastAST as a practical solution for real-time, resource-efficient audio analysis applications. Moreover, the customizable reduction factor allows for fine-tuning according to specific speed and accuracy requirements. Overall, FastAST represents a significant step forward in advancing the efficiency and effectiveness of audio classification systems.

\clearpage

\bibliographystyle{IEEEtran}
\bibliography{mybib}

\end{document}